\documentclass[conference]{IEEEtran}
\IEEEoverridecommandlockouts
\usepackage{cite}
\usepackage{amsmath,amssymb,amsfonts}
\usepackage{algorithmic}
\usepackage{graphicx}
\usepackage{textcomp}
\usepackage{xcolor}
\def\BibTeX{{\rm B\kern-.05em{\sc i\kern-.025em b}\kern-.08em
    T\kern-.1667em\lower.7ex\hbox{E}\kern-.125emX}}
\begin{document}

\title{New approach to MPI program execution time prediction\\
\thanks{This work is supported by Russian Ministry of Science and Higher Education, grant N 05.613.21.0088, unique ID RFMEFI61318X0088 and the National Key R\&D Program of China (2017YFE0123600)}
}

\author{\IEEEauthorblockN{Andrey Chupakhin}
\IEEEauthorblockA{\textit{Lomonosov Moscow State University}\\
Moscow, Russian Federation\\
andrewchup@lvk.cs.msu.ru}
\and
\IEEEauthorblockN{Alexey Kolosov}
\IEEEauthorblockA{\textit{Lomonosov Moscow State University}\\
Moscow, Russian Federation\\
akolosov@cs.msu.ru}
\and
\IEEEauthorblockN{Ruslan Smeliansky}
\IEEEauthorblockA{\textit{Applied Research Center for} \\
\textit{Computer Networks (ARCCN)} \\
\textit{Lomonosov Moscow State University}\\
Moscow, Russian Federation \\
smel@cs.msu.ru}
\and
\IEEEauthorblockN{Vitaly Antonenko}
\IEEEauthorblockA{\textit{Applied Research Center for} \\
\textit{Computer Networks (ARCCN)} \\
\textit{Lomonosov Moscow State University}\\
Moscow, Russian Federation \\
anvial@lvk.cs.msu.su}
\and
\IEEEauthorblockN{Gleb Ishelev}
\IEEEauthorblockA{\textit{Lomonosov Moscow State University} \\
Moscow, Russian Federation \\
Gleb.Ishelev@skoltech.ru}
}



\maketitle

\begin{abstract}
The problem of MPI programs execution time prediction on a certain set of computer installations is considered. This problem emerge with orchestration and provisioning a virtual infrastructure in a cloud computing environment over a heterogeneous network of computer installations: supercomputers or clusters of servers (e.g. mini data centers). One of the key criteria for the effectiveness of the cloud computing environment is the time staying by the program inside the environment. This time consists of the waiting time in the queue and the execution time on the selected physical computer installation, to which the computational resource of the virtual infrastructure is dynamically mapped. One of the components of this problem is the estimation of the MPI programs execution time on a certain set of computer installations. This is necessary to determine a proper choice of order and place for program execution. The article proposes two new approaches to the program execution time prediction problem. The first one is based on computer installations grouping based on the Pearson correlation coefficient. The second one is based on vector representations of computer installations and MPI programs, so-called embeddings. The embedding technique is actively used in recommendation systems, such as for goods (Amazon), for articles (Arxiv.org), for videos (YouTube, Netflix). The article shows how the embeddings technique helps to predict  the execution time of a MPI program on a certain set of computer installations.
\end{abstract}


\section{Introduction}
The idea of building a virtual infrastructure in a cloud computing environment over a heterogeneous network of computer installations is not new \cite{GENI}, \cite{GRANIT}, \cite{ORCA}, \cite{GEANT}. To do this several administrative authorities called federates, allocate some their computational resources for the cloud computing environment operation. Such unions, called federations, have already been established \cite{GENI}, \cite{OPHELIA}, \cite{FED4FIRE}, \cite{FED4FIRE+}, some are only planning to be created \cite{FABRIC}, \cite{MC2E_NEC_2019}. Each federate can consists of several computer installations with different computer powers, such as high-performance computing cluster, data center or supercomputer. Federate provides resources of its computing environment \footnote{Further the computing environment will be understood as a heterogeneous network of computer installations} for creating a virtual infrastructure in accordance with the access policy set by the federate administration. Further in this article, the term program will be understood as MPI Program.

One of the main criteria for the effectiveness of such cloud computing environment is the time spending by a program in this environment. This time consists of the time spent by program in the queue (waiting time) and the program execution time (execution time). This value depends on the resources allocation algorithms of the cloud environment (CE) (mapping virtual computer installations to physical ones) and the queue discipline, taking into account the heterogeneity of physical computer installations.

Consider the “path” of a program from the entering the CE up to getting the result of its execution:
\begin{enumerate}
    \item The user forms a set of input data: the text of the program, the program arguments,  program launch arguments (e.g. the number of MPI processes or  the requested resources (cpu, ram)), the input data for program, a special script to compile program;
    
    \item The user send a set of input data to the CE through a single portal of the federation. The input data from the portal come to the orchestrator of the CE, which is responsible for implementing the virtual infrastructure and executing the programs in it;
    
    \item The orchestrator responsible for federation resources allocation and has a unique algorithm to do this. This algorithm determines the best order and the best federate and physical computer installation inside federate for program execution, in sense of the above mentioned effectiveness criterion. When computer installation is selected this program is sent to there;
    
    \item Each federate has its own queue of programs. The arrived program is  processed by the local resources scheduler and will start running at a certain time defined by the local scheduler;
    
    \item When the program execution is completed the results is returned to the orchestrator which in its term sends it in the proper form to the users.
\end{enumerate}

From the path description above, it can be seen that step 3 (meta scheduling) and step 4 (local scheduling) can be optimized in the context of this article. The focus of this article is step 3, namely: how to choose the most effective computer installation that is a part of the federation computing environment for the received program based on the minimum execution time criterion. The article offers algorithms for program execution time prediction on a certain set of computer installations based on the execution history of the program on different computer installations.

The program execution time on the specific computer installation, as well as the waiting time in the queue, can be predicted based on the history of its launches on this computer installation \cite{Gibbons_1997}, \cite{Kapadia_1999}, \cite{Li_2004}, \cite{Mohr_2003}. To do this, various extrapolation algorithms can be used \cite{Iverson_1999}, \cite{Liu_2008}, regression \cite{Ridge_Regression}, or more complex algorithms are used, such as the ensemble of decision trees (Random Forest) \cite{Random_Forest}. The main disadvantage of these algorithms is that they applicable only to a single certain computer installation. But the point of the question of this article is a program execution time prediction on a certain set of computer installations. These are the computer installations whose characteristics meet the requirements of the virtual infrastructure. Of course, the algorithms that were mentioned above can be used to get estimates for programs execution time on the several computer installations. However, this requires a history of running  of this program on each computer installation from the set. As a rule, this information isn’t available.

The logic for choice of computer installation based on the histories of  program execution can be described as follows. One of the well-known algorithms \cite{Gibbons_1997}, \cite{Kapadia_1999}, \cite{Li_2004}, \cite{Mohr_2003} is applied to program execution histories to estimate  program execution time on each of the computer installation from a certain set of computer installations. With obtained estimates, one can either developing scheduling for a group of programs, or follow a greedy strategy and send each program to the computer installation that has the minimum execution time.

As noted above, in order to allocate programs between several computer installations, a history of program executions on each of the computer installations is required. 

As demonstrated in this paper this requirement can be relaxed: to predict the program execution time on several computer installations, some running histories of this program on them are sufficient. Moreover, it is not necessary that each program launches on each computer installation. The accuracy of the prediction depends on the number of program running histories on computer installations from a certain set. To do this, the article proposes a new approaches to estimate program execution time, which allows predicting the program execution time on a certain set of computer installations. The algorithms presented in this article for implementing these approaches allow one to predict the execution time of a program on a certain computer installation, even there is no execution history the program on it.

In the proposed approaches, the problem of the program execution time prediction is reduced to filling empty entries in the matrix ``Programs-Computers'' built for a given set of programs and a given set of computer installations, in the entries which is an execution time a certain program with certain sets of arguments on a specific computer installation. For more information about the mathematical statement of this problem see Section 2.

In this article proposed two approaches to the program execution time prediction problem on a certain set of computer installations. The first one is based on computer installations grouping based on the Pearson correlation coefficient \cite{Pearson}. It is shown that this method should be used in the case of a "densely" (dense matrix) filled matrix "Programs-Computers" (at least $95\%$ of entries are filled). The second one is based on decomposition of “Programs-Computers” matrix into vector representations of computer installations and programs, so-called embeddings. Here embedding is treated as a structured type of unstructured data. For example, the form of a vector implies the structure that is important here in contrast to hashing, which simply converts a potentially infinite object into a finite one. In the paper it is shown how to use embedding of program and embedding of computer installation to predict the program execution time on specific computer installation. To calculate embeddings, it’s proposed to use the technique [Matrix-Decomposition] decomposition  of the matrix "Programs-Computers". In Section 6 it is  shown by experimental studies that the second method works well with sparse matrices. Histories of runnings of MPI and OpenMP benchmarks \cite{MPI_LOGS}, \cite{ACCEL_OMP} on a dozens of different computer installations were used for experiments. This data is described in Section 4.

It’s important to emphases that, as it will be shown below, the proposed decomposition technique results embeddings of dimension 1, that allows to place a total order relation on a set of computer installations and programs.

The rest part of the article is structured as follows. The 5th Section contains a description of the developed algorithms for programs execution time prediction: an algorithm based on computer installations grouping that works well for dense matrices and an algorithm for obtaining vector representations (embeddings) of computer installations and programs that works well for sparse matrices. Section 7 describes the application of the developed algorithms to solve the problem of "minimization of the time spent by the program inside the cloud environment". In the last 8-th section, conclusions are presented and goals for further research are formulated.

\section{Problem Description}
We consider the case when almost nothing is known about the program and the computer installation. The program source code and binary code  are not available, the architecture of the computer installation is unknown. Only the program execution time, its arguments (see point 1 in Section 1) and the numbers of used resources of computer installation are known.

Further, the program will be considered in the form of job which has the following characteristics: the time when job is submitted to the queue, the start time and end time, resource utilization (cpu, ram, disk), input data, arguments of the program and arguments of the computing environment in which this job is started (e.g., the number of MPI processes, the number of threads for OpenMP, and the amount of resources needed to complete the job).

It is assumed that only the total amount of resources on the computer installation is known: the number of nodes\slash CPUs\slash cores, the amount of RAM and the storage space.

We will use the following notations to formulate the problem:
\begin{enumerate}
    \item $P_i$ -- unique identifier of  program;

    \item $\{P_1, P_2, \ldots , P_N\} = \{P_i\}$ -- a set of $N$ unique identifiers of programs; $Arg$ -- arguments of program: program’s arguments (for the executable file) and arguments of the computing environment where the program is running (the number of MPI processes, the amount of requested computing nodes \slash CPUs \slash cores);
    
    \item $\{Arg_1, Arg_2, \ldots , Arg_{A_i}\} = \{Arg_j\}$ -- here each $Arg_i$ is the set of arguments of $i$-th program. A total amount of different sets of arguments -- $A_i$;
    
    \item $[(P_i, Arg_1), (P_i, Arg_2), \ldots , (P_i, Arg_{A_i}), (P_i, Arg_1)]$ -- a history of $P_i$ program running, arguments can be the same in some pairs;
    
    \item $C_i$ -- unique identifier of computer installation;
    
    \item $\{C_1, C_2, \ldots, C_M\} = \{C_i\}$ -- a set of $M$ unique identifiers of computer installations;
    
    \item Matrix $PC$ ($P$ -- ``program'', $C$ -- ``computer'') -- the matrix where rows correspond to history of running of programs, columns correspond to the computer installations. This matrix is divided into groups of rows. There are $N$ groups in total (by the number of the programs). The $i$-th group corresponds to the history of running of the $i$-th program. The group consists of $A_i$ rows, where each matrix entry display the execution time of the program $P_i$ with the corresponding arguments $Arg_i$ on the corresponding computer installation. The matrix entry is empty if the program was never run on this computer installation.
\end{enumerate} 

\begin{table}[htbp]
\caption{Example of $PC$ matrix}
\begin{center}
\begin{tabular}{|c|c|c|c|c|}
\hline
\textbf{$P/C$} & \textbf{$C_1$} & \textbf{$C_2$} & \textbf{$C_3$} & \textbf{$C_4$} \\
\hline
$P_1, Arg_1$ & ? &  ? & 90 & ?  \\
\hline
\ldots & \ldots &  \ldots & \ldots & \ldots  \\
\hline
$P_1, Arg_{A_1}$ & 5 &  10 & 15 & 25  \\
\hline
\ldots & \ldots &  \ldots & \ldots & \ldots  \\
\hline
$P_i, Arg_1$ & 25 &  35 & ? & 56  \\
\hline
\ldots & \ldots &  \ldots & \ldots & \ldots  \\
\hline
$P_i, Arg_{A_i}$ & 60 & 75 & 96 & ?  \\
\hline
\ldots & \ldots &  \ldots & \ldots & \ldots  \\
\hline
$P_N, Arg_1$ & ? &  34 & 67 & ?  \\
\hline
\ldots & \ldots &  \ldots & \ldots & \ldots  \\
\hline
$P_N, Arg_{A_1}$ & 100 & 146 & 245 & 300  \\
\hline
\end{tabular}
\label{PC_Matrix_Example}
\end{center}
\end{table}

In the Table~\ref{PC_Matrix_Example} an example of a $PC$ matrix is presented. Empty entries are marked with a question mark. The $PC$ matrix shows the history of finished program executions. It can be built from log files of the computer installations -- $C_1,..., C_M$.

The terms of the notations introduced above the problem of program execution time prediction on a set of computer installations can be formulated as following:
\begin{itemize}
    \item \textit{Given}
    \begin{itemize}
        \item A set of $N$ unique identifiers of programs -- $\{P_i\}$;
        \item For each program $P_i$ a set with $A_i$ sets of arguments -- $\{Arg_j\}$, $|\{Arg_j\}| = A_i$;
        \item A set of $M$ unique identifiers of computer installations -- $\{C_i\}$;
        \item A $PC$ matrix where entries contain the execution times of programs $P$ (with the corresponding arguments -- $Arg$) on computer installation $C$;
    \end{itemize}
    
    \item \textit{To find}
    \begin{itemize}
        \item Fill in empty entries of the $PC$ matrix, or in other words predict the program execution times on the computer installations.
    \end{itemize}
    
\end{itemize}

To evaluate the program execution time prediction, the prediction error is calculated as follows:
\begin{equation}
Prediction Error(P_i, Arg_j) = \frac{|predict - target|}{target}\label{error}
\end{equation}

where \textit{predict} is predicted time of program $P_i$ with arguments $Arg_j$, \textit{target} is a true execution time of program $P_i$ with arguments $Arg_j$.

The \textit{total prediction error} is calculated as an average of the errors calculated using the equation \eqref{error} for all programs. This allows you to compare algorithms among themselves.

\section{Related Works}
The problem of the execution time prediction has been studied for a very long time. For example, for real-time systems, the problem of estimation the worst-case execution time (WCET) is still one of the main problems, since WCET programs are used to build a schedule of tasks in real-time systems. In \cite{tese_Marcos_FINAL}, the execution time of specific applications on GPUs is predicted. In \cite{Snavely_2002} a model of the program and the computer are built in order to estimate the program execution time on the computer. In \cite{Nadeem_2017} the execution time of programs on Grid systems are predicted. For time prediction purposes are used Analytical approaches \cite{Brehm_1997}, \cite{Bacigalupo_2005}, \cite{Yero_2006}, statistical approaches \cite{Barnes_2008}, \cite{Wu_2011}, \cite{Wu_2006} approaches based on historical data \cite{Gibbons_1997}, \cite{Kapadia_1999}, \cite{Li_2004}, \cite{Mohr_2003}, time series prediction \cite{Iverson_1999}, \cite{Liu_2008}, neural networks are also used \cite{Ipek_2005}, \cite{Lee_2007}. Execution time can be predicted from test runs \cite{Yang_2005}.

All algorithms for predicting program execution time mentioned above use the history of program executions to make a prediction. Their main drawback is that these algorithms only work when the history of program executions is known on the computer installation, i.e. to predict the execution time on a certain computer installation, you need more than one run of the program on this computer installation because the history of executions have to consists of at least two executions.  
    
Besides above, in order to build a program model or make test runs, you need the source code and executable files, and this information is often not available. In addition, almost all articles that consider the prediction of the program execution time often considered only for one computer installation. However, in the point of question of this paper it has to be able to estimate the program execution time on a set of computer installations in that case when there is a minimum amount of information about computer installations and programs, actually available.

\section{Data for Prediction Model}
Information about programs (benchmarks) executions on many computer installations from the website \textit{spec.org} was used for the experiments presented in this article. This information was used to form the matrix ``Programs-Computers'' $PC$, described in Section 2 above. Two data sets with the execution results of programs on different computer installations are presented in \cite{MPI_LOGS}:
\begin{enumerate}
    \item MPIL2007 -- $12$ programs, $163$ computer  installations;
    \item MPIM2007 -- $13$ programs, $396$ computer  installations.
\end{enumerate}

It should be taken into account that the data on the website \textit{spec.org} is constantly updated. The data are used in the presented research dated \textit{April 8, 2020 12:13}. The history of changes can be viewed in \cite{MPI_LOGS_HISTORY}. The execution results of OpenMP benchmarks running \cite{ACCEL_OMP} were also used as a data to form the $PC$ matrix.

The problem formulated in Section 2 implies that the program can be run with different arguments. However for the simplicity reason we will assume that each program was run only with one set of arguments, i.e. one program corresponds to one row in the $PC$ matrix. Note that this assumption does not limit the generality of the proposed approaches. 

A brief description of the data sets is provided in the Table~\ref{Data_Sets_Description}. Each entry in \cite{MPI_LOGS} was considered as a separate computer installation.

\begin{table}[htbp]
\caption{Results of MPI and OpenMP programs launches on various computer installations}
\begin{center}
\begin{tabular}{|c|c|c|c|c|}
\hline
 & \textbf{Number of} & \textbf{Number of} & \textbf{Benchmark} &  \\
\textbf{Name} & \textbf{computer} & \textbf{programs} & \textbf{type} & \textbf{Link to data} \\
 & \textbf{installations} &  &  &  \\
\hline
MPIL2007 & 163 & 12 & MPI & \cite{MPI_LOGS}  \\
\hline
MPIM2007 & 396 & 13 & MPI & \cite{MPI_LOGS}  \\
\hline
ACCEL\_OMP & 25 & 15 & OpenMP & \cite{ACCEL_OMP}  \\
\hline
\end{tabular}
\label{Data_Sets_Description}
\end{center}
\end{table}

The used data was uploaded on Github\footnote{https://github.com/andxeg/CloudCom\_2020\_Embeddings} \cite{Our_Code_and_Data}.

\section{Proposed Solutions}
To solve the problem of the programs execution time prediction (see Section 2) the following algorithm was developed (for brevity, only the algorithm scheme is presented):
\begin{enumerate}
    \item \textit{Data preparation}. Form a $PC$ matrix of dimension $N$ by $M$;
    \item \textit{Execution time prediction}. Applying one of the algorithms described below:
    \begin{enumerate}
        \item \textit{Ridge}. Using ridge regression \cite{Ridge_Regression} for each row of the $PC$ matrix. In other words the unknown execution time on some computer installation was predicted from the known execution times of the particular program on all other computer installations;
        \item \textit{Cliques}. Grouping computer installations using the Pearson correlation coefficient;
        \item \textit{Decomposition}. Apply matrix decomposition \cite{Matrix_Decomposition} or tensor decomposition \cite{TENSOR_DECOMPOSITION} of the $PC$ matrix in order to fill in empty entries of the $PC$ matrix;
        \item Apply an ensemble of algorithms: \textit{Ridge} + \textit{Cliques} + \textit{Decomposition};
    \end{enumerate}
    \item \textit{Evaluation of prediction quality}. The prediction of program execution times is evaluated as follows:
    \begin{enumerate}
        \item For each program a prediction error is calculated by the equation \eqref{error} specified in Section 2;
        \item The error average is a general prediction error.
    \end{enumerate}
\end{enumerate}

As it was mentioned above each program was run only with one set of arguments, i.e. one program corresponds to one row in the $PC$ matrix.  If the program was executed  on multiple sets of arguments, it’s mapped to multiple rows in the $PC$ matrix. In this case, one can first use the history-based execution time prediction algorithm \cite{Gibbons_1997}, \cite{Kapadia_1999}, \cite{Li_2004}, \cite{Mohr_2003} to fill in the empty entries, and then use the algorithm described above in 2a-2d. For more information about the case when a single program corresponds to multiple rows in the $PC$ matrix, see Section 7.

\subsection{Computer installations grouping based on the Pearson correlation}
The Pearson correlation coefficient \cite{Pearson} characterizes the presence of a linear relationship between two sets of numbers. The columns of the $PC$ matrix are considered as such sets. By calculating the correlation between the columns $C_i$ and $C_j$, one can detect a linear relationship between the programs execution times on the corresponding computer installations. If the Pearson correlation between columns $C_i$ and $C_j$ is close to $1$ in modulus, then there is a linear relationship between the execution times of programs on $C_i$ and $C_j$ computer installations. Therefore, a column with program execution times on the $C_i$ computer installation can be obtained by multiplying by a certain constant the column with program execution times on the $C_j$ computer installation.

The grouping procedure for computer installations consists of the following steps:
\begin{enumerate}
    \item Calculate Pearson correlation for each pair of columns of the matrix $PC$ -- total $\frac{M(M-1)}{2}$ pairs, where $M$ is the amount of computer installations;
    
    \item Build the graph of computer installations:
    \begin{enumerate}
        \item Each vertex represents  one computer installation;
        
        \item An edge between the $C_i$ and $C_j$ computer installations exists if the Pearson correlation between the corresponding columns in the $PC$ matrix is modulo greater than some threshold. The value of threshold is an algorithm parameter;
    \end{enumerate}
    
    \item Find all cliques \cite{Cliques} in the graph of computer installations. Cliques searching is NP-complete problem. To solve this problem you can use the algorithm from \cite{Cliques_algo}. To speed up the experiments, a simplified cliques search algorithm was implemented \cite{Our_Code_and_Data};
    
    \item Each clique is a group of computer installations with execution times of each program are  linear related, i.e. each pair of computer installations in such a group has execution times related by a specific scalar coefficient. Pay attention that the same computer installation can fall into several groups;
    
    \item The resulting cliques are groups of computer installations.
\end{enumerate}

To search for cliques, the simplified algorithm was developed with complexity no bigger than $N^3$ \cite{Our_Code_and_Data}, which is significantly less than the complexity of the algorithm for searching for all maximum size cliques \cite{Cliques_algo} – $3^{\frac{N}{3}}$. It finds at least one clique for each vertex in the graph. The disadvantage of the proposed algorithm is that it doesn’t find all the maximum cliques. However, if threshold for the value of Pearson correlation is close to $1$, then further prediction algorithm described below is pretty good even  some vertexes in the cliques would be missed.

Based on computer installations grouping, the procedure for predicting program execution time on the computer installation can be described as follows:
\begin{enumerate}
    \item Select $P$ -- the program for execution time prediction;
    
    \item Select $C$ -- the computer installation where $P$ program will be run;
    
    \item Determine the group that $C$ belongs to;
    
    \item If the computer installation doesn’t belong to any group of computer installations, then make a prediction using the Ridge Regression (See Section 5, point 2.a). Regression is applied for the entire row in $PC$ matrix correspond to the $P$ program;
    
    \item If the computer installation $C$ belongs to a certain group, then there is a clique in the graph of computer installations, in which all the computer installations, including $C$, are connected to each other. This means that Pearson correlation is greater than some given threshold for each pair \textit{(computer installation $C$; some other computer installation from the clique -- $C'$)}. Therefore, it’s possible to calculate the coupling coefficient between the programs execution times for two computer installations from a pair;
    
    \item Then multiply the $P$ program execution time on the computer installation $C'$ on this coefficient. As result we get the estimate of $P$ program execution time on the considered computer installation $C$;
    
    \item This procedure should be applied to each computer installation from the clique, and then calculate the average of the  execution times. The  average time is the estimation of the $P$ program execution time on the $C$;
    
    \item If it isn’t possible to calculate Pearson correlation (e.g, the considered program $P$ hasn’t been run on any of the computer installations from clique), but corresponding row in $PC$ matrix for $P$ program is  non-empty then one need to use Ridge Regression for the prediction. See \textit{Step 4}. If the row for $P$ program is empty then method described in Section 7 is used.
\end{enumerate}

The error of prediction for the algorithm presented above can be estimated by the equation \eqref{error} from Section 2.

\subsection{Matrix decomposition and Tensor decomposition}
The problem of programs execution time prediction is very similar to the problem solved in the recommendation systems. In these systems, there are usually two types of objects and a relationship between them that can be quantified. For example, objects can be users and movies, users and books, users and goods, and so on. The relationship between them is often a rating that the user puts on a movies, books, or goods. You can build a rating matrix where the rows (or columns) correspond to movies, books, or goods, and the columns (or rows) correspond to users. This matrix is often sparse, since there are a lot of users and objects, and users can't physically evaluate all the objects. In the field of recommendation systems, there is a problem is to determine the ratings of all users for all objects, in other words, you need to fill in the empty entries in the rating matrix.

Consider using the following analogy: users are computer installations, films, books or products are programs, and user ratings are execution times. Thus, computer installations ``evaluate'' programs and the smaller the rating (execution time), the better.

The problem of filling empty entries in recommendation systems is solved by matrix decomposition method \cite{Matrix_Decomposition}. We propose to use the matrix decomposition technique to solve the problem of programs execution time prediction. 

The result of application matrix decomposition techniques to some matrix is two or more matrices, the product of which gives a matrix that approximates the original one. For empty entries of the original matrix, i.e. for unknown values, the product of the matrices gives values that estimate the unknown values.

$PC$ matrix decomposition allows one to get a vector representations of programs and computer installations, which have a remarkable property: the scalar product of the vector representation of the program and the vector representation of the computer installation is the program execution time on the computer installation. Thus, the program execution time can be divided into two components: one is related to the program itself, the other to the computer installation.

The vector representations of programs and computer installations is embeddings \cite{EMBEDDINGS} of programs and computer installations. Embedding techniques and methods of applying embeddings are very well known in such areas as NLP \cite{NLP}, topic modeling \cite{TOPIC_MOD}, recommendation systems \cite{RECOM_SYS}.

\begin{figure}[htbp]
\centering
\includegraphics[width=3.3in]{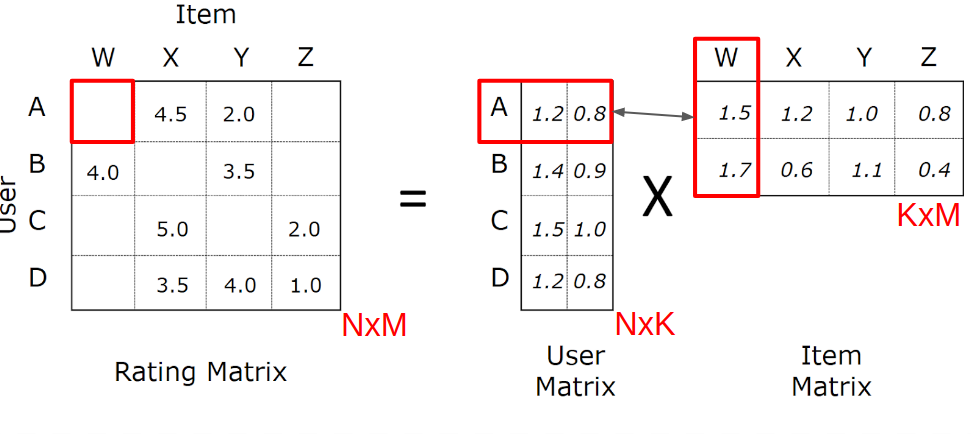}
\caption{Example of matrix decomposition.}
\label{Matrix_Decomposition}
\end{figure}

The example of matrix decomposition is shown in Fig.~\ref{Matrix_Decomposition}. In the \textit{Rating Matrix} rows correspond to \textit{Users}, and columns correspond to objects of ratings -- \textit{Items}. Each entry in the matrix is either empty or the user's rating. $K$ is a matrix decomposition parameter that can be changed to achieve the required accuracy.

The result of matrix decomposition of \textit{Rating Matrix} of dimension $N$ by $M$ is two matrices: \textit{User Matrix} of dimension $N$ by $K$ and \textit{Item Matrix} of dimension $K$ by $M$. The \textit{User Matrix} contains a vector representation of \textit{users} in rows, and the \textit{Item Matrix} contains a vector representation of \textit{items} in columns. The product of the user's vector representation and the item's vector representation is the user's rating for this item. The $K$ parameter affects not only the accuracy, but also the number of components in the vector representation.

The following matrix decomposition algorithms were used to solve the problem of programs execution time prediction: Singular Value Decomposition (SVD) \cite{SVD}, Alternating Least Square (ALS) \cite{ALS_DECOMPOSITION}. It was also used the tensor decomposition algorithm \cite{TENSOR_DECOMPOSITION}, based on adaptation of  SVD decomposition for tensors . For tensor decomposition, a $3$-dimensional tensor should be formed from the original $PC$ matrix. First, the time axis was divided into time slots of a certain size. Next, the tensor is formed from several $PC'$  matrices (the rows – programs, the columns – computer installations), each of which is associated with a certain time slot. The entries of the $PC'$ matrices contain $1$ if the program execution time on the computer installation hits the corresponding time slot, and $0$ otherwise.

As a result of experimental evaluation, the ALS algorithm was selected as one with the minimum error (see equation \eqref{error} from Section 2.).

\subsection{Ensemble of algorithms}
In Sections 5.1 and 5.2, algorithms for the program execution time prediction on a certain set of computer installations were proposed. These algorithms can be combined into an ensemble of algorithms to improve the accuracy of the prediction \cite{Ensemble}. In the article averaging ensemble \cite{Ensemble} is used, where the estimation for execution time calculated for each program, as the average value of the execution time estimations of the following algorithms:

\begin{enumerate}
    \item Ridge Regression -- \textit{Ridge};
    \item Computer installations grouping based on the Pearson correlation -- \textit{Cliques};
    \item Matrix decomposition by the ALS algorithm -- \textit{Decomposition}.
\end{enumerate}

\section{Experiments}
In this section the results of experimental studies of the algorithms proposed in Section 5 are presented.

\textit{The purposes of the experiments are analysis of the quality of prediction results:}

\begin{enumerate}
    \item based on grouping computer installations by the  Pearson correlation;
    \item based on ALS matrix decomposition algorithm. Selecting the parameter $K$ – the number of components in the vector representation of programs and computer installations (see Section 5.2);
    \item by the ensemble of algorithms;
    \item by the proposed algorithms with outliers in the source data.
\end{enumerate}

\textit{Data}

The data sets described in Section 4 were used in experiments. These data sets were used to form the $PC$ matrix.

\textit{The experimental technique}

Implemented algorithms \cite{Our_Code_and_Data} were run with different data sets during experimentation. 

\subsection{Analysis of the quality of prediction based on grouping computer installations by  Pearson correlation}
Data set MPIM2007 (see Section 4) with $13$ programs and $396$ computer installations were used for the experiments.

The algorithm based on grouping computer installations by Pearson correlation is very sensitive to the presence of outliers in the data, as well as to what extend the $PC$ matrix is low-density. In order to analyze  the quality of prediction by this algorithm for each entry of the $PC$ matrix, an execution time prediction was made using information from the remaining entries of this matrix. Thus, only one value was removed from the matrix, and then the algorithm described in Section 5.1 was applied. Results are presented in the Table~\ref{Pearson_Results}.

\begin{table}[htbp]
\caption{Results of the prediction  based on grouping computer installations by Pearson correlation}
\begin{center}
\begin{tabular}{|c|c|c|c|c|c|}
\hline
 & & & \textbf{Groups with} & \textbf{Prediction} & \textbf{Average} \\
\textbf{Exp.} & \textbf{Corr.} & \textbf{Groups} & \textbf{size 1} & & \textbf{Error} \\
\hline
1 & - & - & - & Regression & 0.25 \\
\hline
2 & 0.97 & 46 & 27 & In groups &  0.068 \\
\hline
3 & 0.97 & 46 & 27 & In groups + & 0.115 \\
 & & & & Regression & \\
\hline
\end{tabular}
\label{Pearson_Results}
\end{center}
\end{table}

3 experiments (Exp.) were made by following steps:
\begin{enumerate}
    \item \textit{Correlation} (Corr.) is the threshold for Pearson correlation coefficient module. It’s necessary to build a graph of computer installations, see Section 5.1;
    
    \item \textit{Number of groups} (Groups) is the number of cliques in graph of computer installations;
    
    \item \textit{Number of groups with size 1} (Groups with size 1) is the number of cliques only with one computer installation;
    
    \item  \textit{Prediction} -- ``Regression'' -- prediction a value in removed entry by applying of Ridge Regression to remaining elements in row; ``In groups'' -- prediction in groups according to the algorithm described in Section 5.1, groups with only one computer installation aren’t considered in this case; ``In groups + Regression'' -- prediction execution time in groups with size more than $2$, to other computer installations used Ridge Regression;
    \item  \textit{Average error} -- after each removing a value from the matrix entry, a prediction is made, and the prediction error is calculated by equation \eqref{error} from Section 2. After that for all entries the arithmetic mean of prediction errors for all entries of the matrix is calculated.

\end{enumerate}

According to the results presented in Table~\ref{Pearson_Results}, the algorithm for programs execution time prediction based on grouping of computer installation by Pearson correlation gives a prediction error on dense matrices of 11.5\% (only single empty entry in PC matrix). Also the results in Table~\ref{Pearson_Results} shows that the number of groups with size $1$ is quite large, therefore the prediction error with and without them differs almost twice: $11.5\%$ and $6.8\%$ respectively.

\subsection{Analysis of the quality of prediction based on ALS matrix decomposition algorithm}
\begin{figure*}[!t]
\centering
\includegraphics[width=\textwidth,height=6cm]{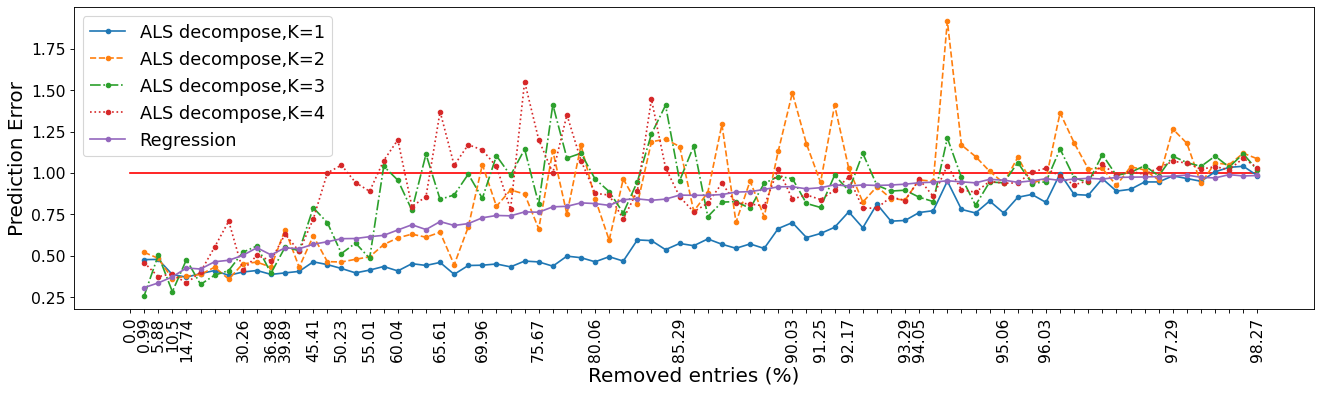}
\caption{Ridge regression and ALS matrix decomposition with K=1,2,3,4.}
\label{Ridge_ALS_graph}
\end{figure*}

Data set MPIM2007 (see Section 4) with $13$ programs and $396$ computer installations used for the experiments.

To study the quality of prediction based on ALS matrix decomposition algorithm $4$ experiments were made: ALS matrix decomposition of $PC$ matrix with $K=1,2,3,4$. The results of the decomposition were compared with each other, as well as with Ridge regression algorithm that was chosen as the basic prediction algorithm.

The Fig.~\ref{Ridge_ALS_graph} contains graphs of the relationship between the prediction error and the percentage of empty entry in the $PC$ matrix for Ridge regression and ALS algorithm with different $K=1,2,3,4$. Prediction error for the algorithms was calculated using the equation \eqref{error} from Section 2. In Fig.~\ref{Ridge_ALS_graph} $X$-axis is the percentage of empty entries in the $PC$ matrix (which randomly was removed from it), $Y$-axis is the prediction error. In the following figures the axises has the same meaning. Some values from $X$-axis were removed for readability.

According to the graphs in the Fig.~\ref{Ridge_ALS_graph} and results from Table~\ref{Pearson_Results} one can make a conclusion that the Ridge and Cliques algorithms work well on dense matrices (up to $1\%$ percentage of empty entries). According to the Fig.~\ref{Ridge_ALS_graph} using the matrix decomposition technique (by comparison with Ridge and Cliques) with $K=1$ gives the best results for sparse matrices in which the number of empty entries is more than $15\%$. Graph for matrix decomposition with $K=1$ is the lowest in the Fig.~\ref{Ridge_ALS_graph}. Even if $80\%$ of the entries are removed from the $PC$ matrix, it can be filled in so that the execution time will deviate by no more than $60\%$ from its true value.

Thus, as a result of experiments, we can conclude that the technique of matrix decomposition with the parameter $K=1$ gives the best solutions when the percentage of empty entries in the $PC$ matrix is more than $15\%$.

An important advantage of usage the matrix decomposition technique is the following. The result of applying this technique is vector representations (embeddings) of programs and computer installations. These embeddings have dimension $1$! This means that you can enter total ordering on the performance on a set of computer installations. The less embedding a particular computer installation has, the less time the program will run on it. This follows from the fact that for $K=1$, the program execution time on a specific computer installation is calculated as the product of two numbers: the program embedding and the embedding of the computer installation.
    
\subsection{Analysis of  the prediction quality of an ensemble of algorithms}
The ensemble of algorithms is described in Section 5.3. The ensemble includes the following algorithms: \textit{Ridge}, \textit{Cliques} and \textit{ALS}. The simple arithmetic mean of the prediction results of the \textit{Ridge}, \textit{Cliques} and \textit{ALS} algorithms was considered. All three data sets – MPIL2007, MPIM2007 and ACCEL\_OMP (see Section 4) were used for the experiments.

The figures contain graphs of the dependence of the prediction error on the percentage of empty entries in PC matrix for Ridge regression, Cliques,  ALS with K=1, and an ensemble of algorithms.

\begin{figure}[!t]
\centering
\includegraphics[width=3.3in]{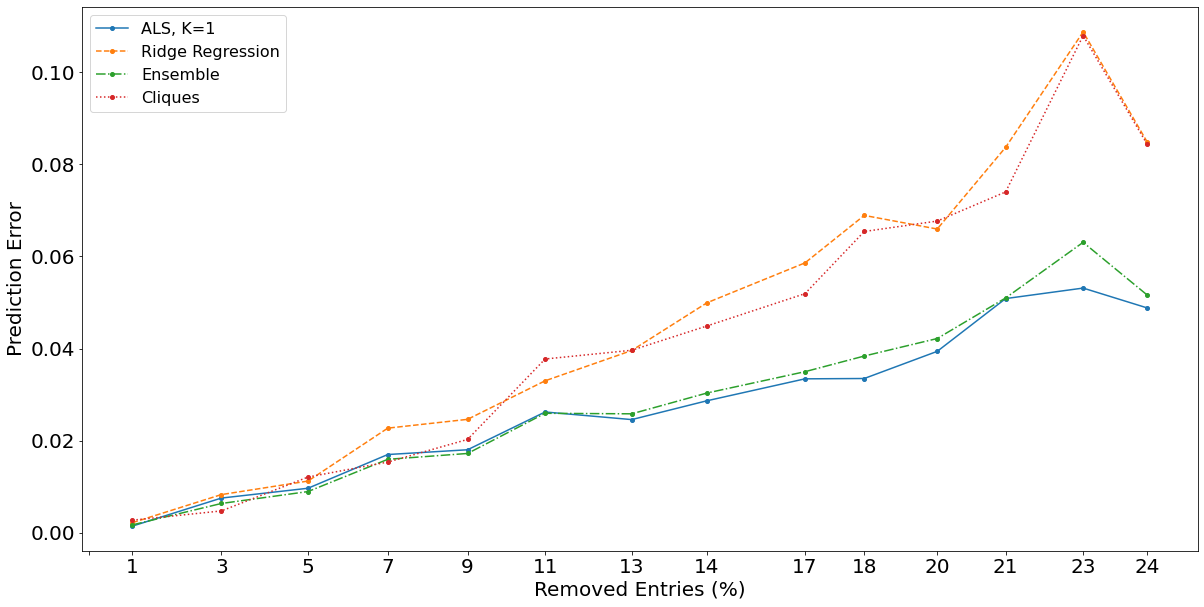}
\caption{Results of Ridge, Cliques, Decomposition and an ensemble of algorithms on MPIL2007 ($1\%$-$24\%$).}
\label{ALL_algo_mid_data_set}
\end{figure}

The Fig.~\ref{ALL_algo_mid_data_set} shows the results for the percentage of empty entries from $1\%$ to $25\%$. The Fig.~\ref{ALL_algo_mid_data_set} shows that the ensemble and Cliques gives the best result when the percentage of empty entries is up to $4\%$. When this percentage is increased, the ALS algorithm gives the best result.

\begin{figure}[!t]
\centering
\includegraphics[width=3.3in]{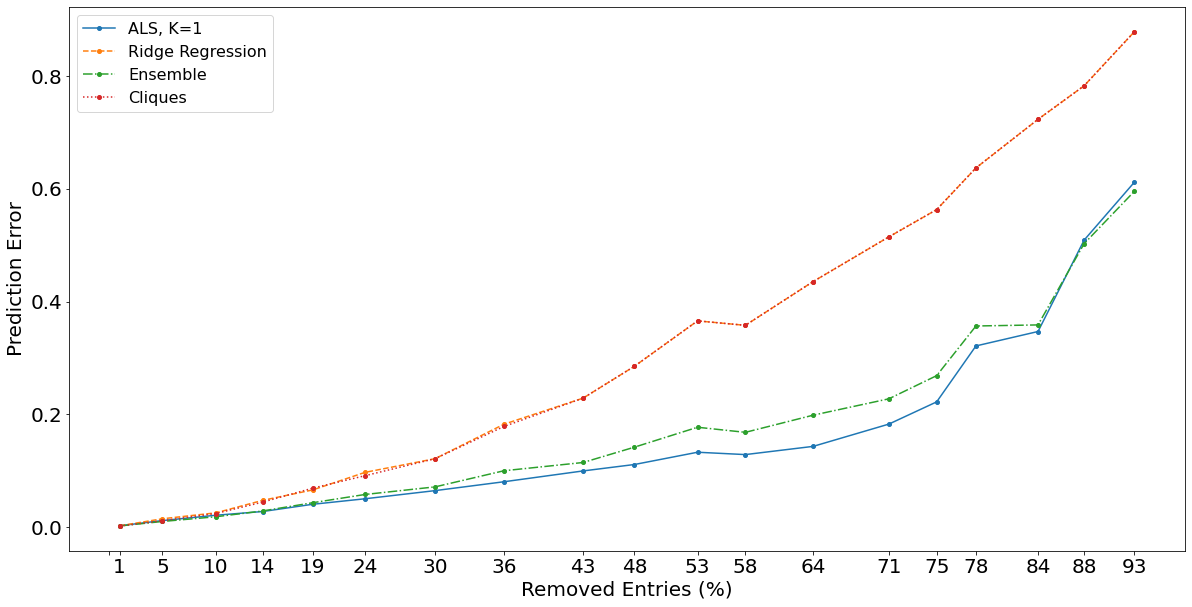}
\caption{Results of Ridge, Cliques, Decomposition and an ensemble of algorithms on MPIL2007 ($1\%$-$94\%$).}
\label{ALL_algo_mid_data_set_full}
\end{figure}

The Fig.~\ref{ALL_algo_mid_data_set} shows the results for the percentage of empty entries from $1\%$ to $94\%$. The Fig.~\ref{ALL_algo_mid_data_set} shows that the ALS algorithm gives the best results in all cases when percentage of empty entries in $PC$ matrix is more than $14\%$.

\begin{figure}[!t]
\centering
\includegraphics[width=3.3in]{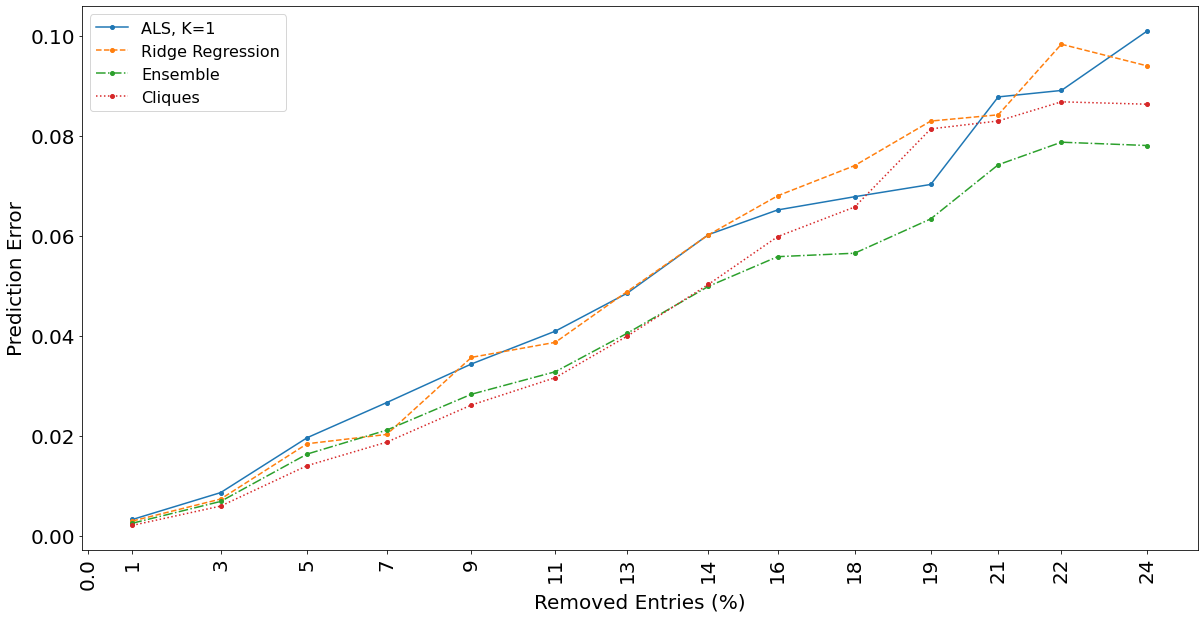}
\caption{Results of Ridge, Cliques, Decomposition and an ensemble of algorithms on MPIM2007 ($1\%$-$25\%$).}
\label{ALL_algo_huge_data_set}
\end{figure}

In Fig.~\ref{ALL_algo_huge_data_set} $X$-axis contains percentage of empty entries in PC matrix from $1\%$ to $24\%$. The Fig.~\ref{ALL_algo_huge_data_set} shows that Ridge regression and Cliques algorithms give the best results up to $14\%$, and the ensemble of algorithms gives the best results after $14\%$.

\begin{figure}[!t]
\centering
\includegraphics[width=3.3in]{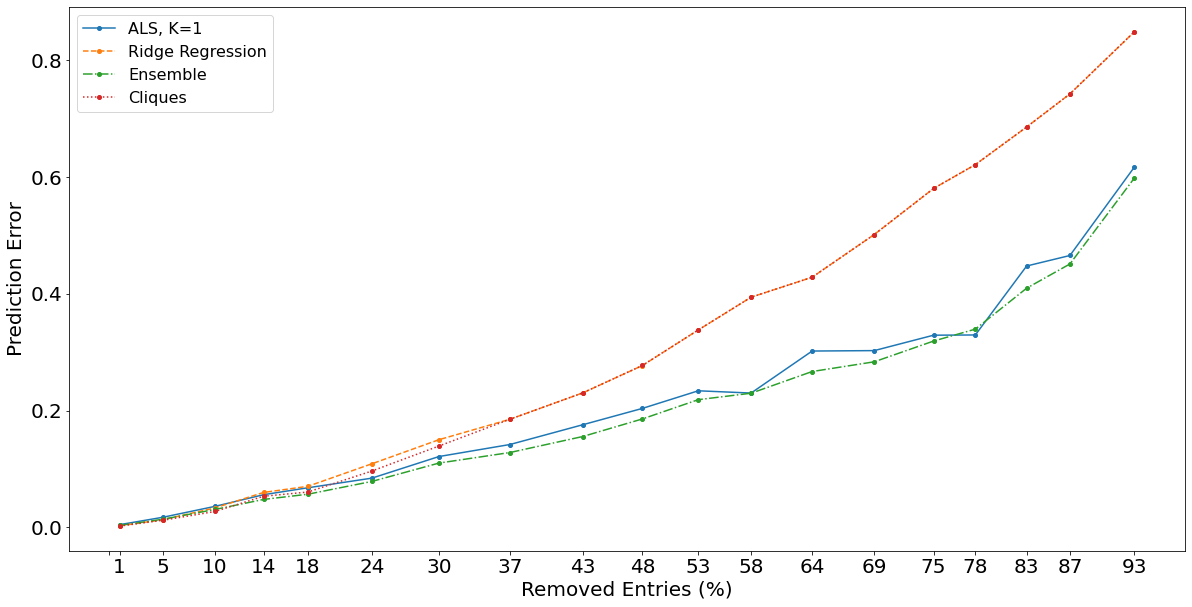}
\caption{Results of Ridge, Cliques, Decomposition and an ensemble of algorithms on MPIM2007 ($1\%$-$94\%$).}
\label{ALL_algo_huge_data_set_full}
\end{figure}

In Fig.~\ref{ALL_algo_huge_data_set_full} $X$-axis contains percentage of empty entries in PC matrix from $1\%$ to $94\%$. The Fig.~\ref{ALL_algo_huge_data_set_full} shows that ensemble of algorithms gives the best results after $14\%$ up to $94\%$.

To test the ensemble of algorithms another experiments was conducted on the data set ACCEL\_OMP with $15$ programs and $25$ computer installations.

\begin{figure}[!t]
\centering
\includegraphics[width=3.3in]{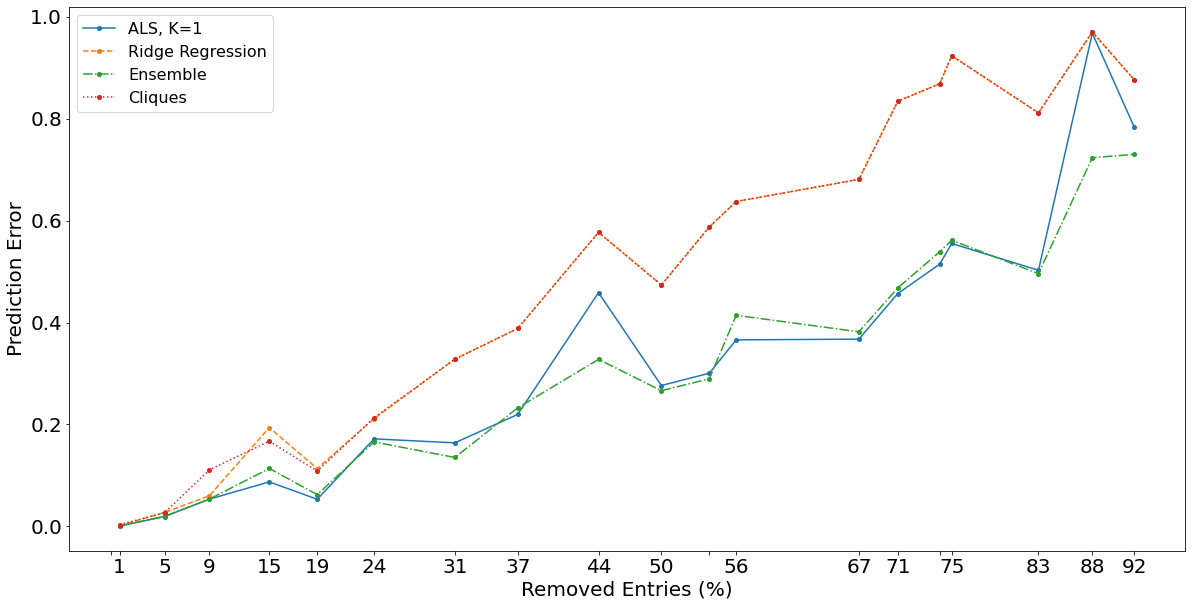}
\caption{Results of Ridge, Cliques, Decomposition and an ensemble of algorithms on ACCEL\_OMP ($1\%$-$92\%$).}
\label{ALL_algo_small_data_set}
\end{figure}

Due to small size of $PC$ matrix – $15$ programs, $25$ computer installations – different algorithms give better results for different percentage of empty entries in PC matrix. It’s difficult to distinguish the dominance of any one algorithm. The Fig.~\ref{ALL_algo_small_data_set} shows that first the best result is given by ALS (up to $19\%$), then ensemble (from $19\%$ to $56\%$), then ALS (from $56\%$ to $83\%$), then the ensemble again (from $83\%$ to $92\%$).

\subsection{Analysis of the prediction quality of the proposed algorithms in the presence of outliers in the source data}
The algorithm for program execution time prediction based on grouping computer installations based on Pearson correlation, like any kind of correlation, is sensitive to outliers in the source data. Therefore, two more experiments were conducted to test the stability of each algorithm to outliers.

For the experiments, the PC matrix was formed, after that $10\%$ of its entries were randomly selected and multiplied:
\begin{itemize}
    \item In the first experiment on a random number in the interval $(0, 4)$;

    \item In the second experiment on a random number in the interval $(0, 10)$.
\end{itemize}

\begin{figure}[!t]
\centering
\includegraphics[width=3.3in]{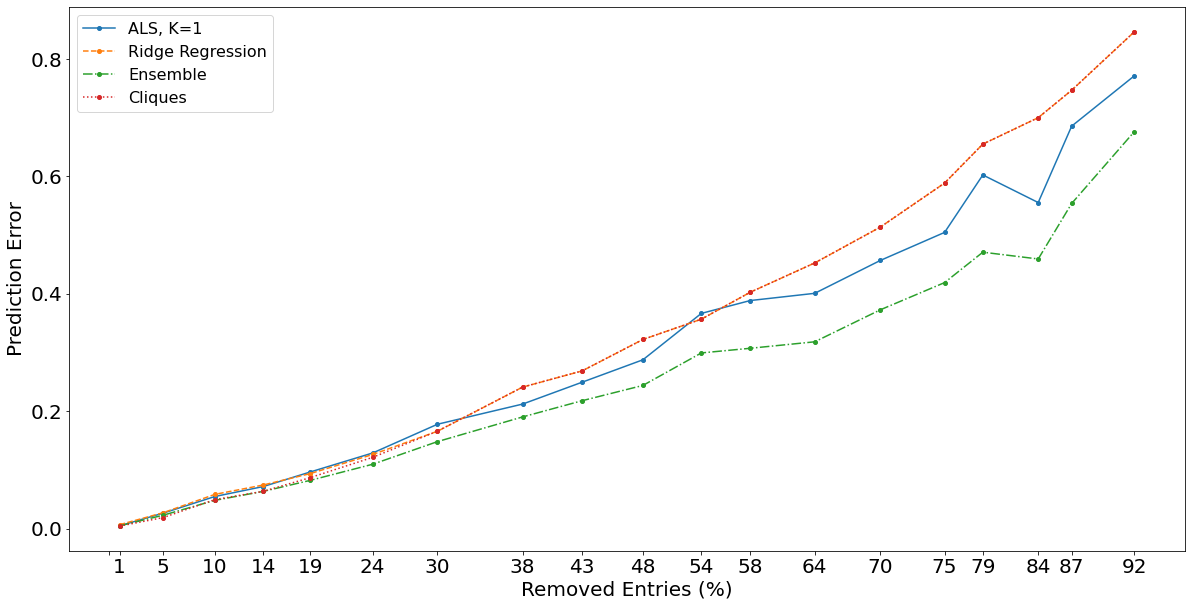}
\caption{Results of Ridge, Cliques, Decomposition and an ensemble of algorithms on MPIM2007 ($1\%$-$92\%$), $10\%$ of outliers, coefficient from $0$ to $4$.}
\label{Outliers_1}
\end{figure}

\begin{figure}[!t]
\centering
\includegraphics[width=3.3in]{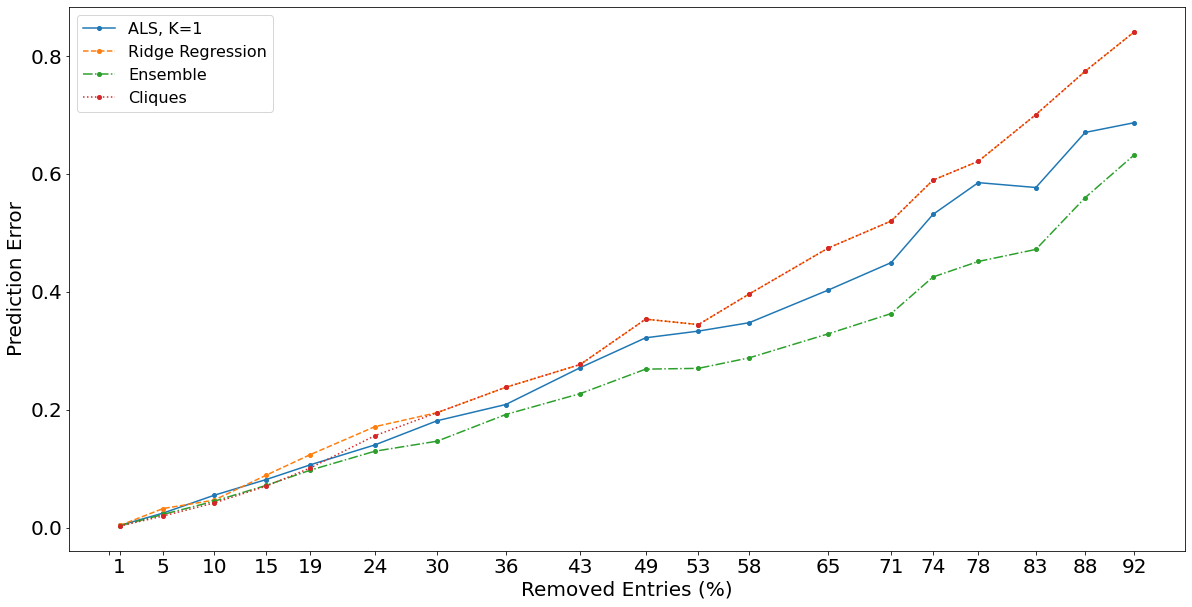}
\caption{Results of Ridge, Cliques, Decomposition and an ensemble of algorithms on MPIM2007 ($1\%$-$92\%$), $10\%$ of outliers, coefficient from $0$ to $10$.}
\label{Outliers_2}
\end{figure}

The Fig.~\ref{Outliers_1} and Fig.~\ref{Outliers_2} shows the results of a prediction based on data with outliers. $X$-axis contains percentage of empty entries in PC matrix. $Y$-axis contains prediction errors. As you can see from the Fig.~\ref{Outliers_1} and Fig.~\ref{Outliers_2} the ensemble of algorithms is almost always better than all other algorithms. The Cliques algorithm also gives a good result up to $19\%$ of the percentage of empty entries.

\subsection{Conclusions from the experiments}
Experiments in the Sections 6.1-6.4 showed that an ensemble of algorithms with a small percentage of empty entries in ``Program-Computer'' matrix up to 4\% and middle percentage of empty entries -- from 19\% to 56\% -- makes a better prediction compared to all other algorithms. Also, as experiment in Section 6.4 showed, the ensemble and ALS show good results even in the presence of outliers in the source data sets. When the percentage of empty entries is much more than 50\%, the best prediction is often made by the ALS algorithm.

Thus, for dense matrices, it is better to use an ensemble of algorithms, for sparse ones -- matrix decomposition, in particular the ALS algorithm.

\section{Execution time prediction application}
Based on the research presented above the following algorithm was constructed for solving the problem of efficient resource allocation in a cloud computing environment over a heterogeneous network of computers:
\begin{enumerate}
    \item Create a $PC$ matrix based on the history of finished program executions;
    \item Fill in empty entries in $PC$ matrix using one of the algorithms described in the article (see Section 5);
    \item Knowing the estimated programs execution times on computer installations apply one of the algorithms below:
    \begin{enumerate}
        \item \textit{Greedy algorithm}. Send the current program to the computer installation where it has the minimum execution time. If this program hasn’t been executed anywhere else, it can be placed on the most powerful computer installation. Performance can be determined by the embedding of the computer installation (see Section 5);
        \item \textit{Build a schedule}. If one plan the calculation for multiple programs, one can create a schedule for running Programs that minimizes the total execution time \cite{Gary_1979}.
    \end{enumerate}
\end{enumerate}

The experiments in Section 6 used ``fixed'' data. A number of programs and a number of computer installations were fixed. But one can not fix them, and change the $PC$ matrix in the process (constantly). So there are two ways to get data for the prediction algorithm:
\begin{enumerate}
    \item Take an arbitrary set of benchmarks from different subject areas and different ``behavior'' with the processor, memory, and network. Start them on all computer installations from the specified set. Build a $PC$ matrix, make a matrix decomposition, and get vector representations of programs and computer installations. Vector representations of computer installations can be used to rank computer installations by inverse ``performance'' if the vector representation has dimension $1$. For new programs that don’t have execution information in the $PC$ matrix, you can predict execution time using only on the ranks of the computer installations;

    \item When choosing a computer installation to start a program, you need to evaluate the program execution time, and not just rely on the ranks of the computer installations. This is important because the behavior of the same program can be completely unpredictable on different computer installations. This is due to the fact that different computer installations can perform certain functions, operations, and commands that are present in the program in different ways. Therefore, to take into account the ``behavior'' of the program, you need to constantly change the matrix: add new rows for new programs, new rows for old programs, but with new arguments, new columns for new computer installations.
\end{enumerate}

Set of benchmarks from the website \textit{spec.org} \cite{SPEC_BENCH} can be used to build a $PC$ matrix.

In the experiments presented in Section 6, each program had only one set of arguments and only one row in the $PC$ matrix corresponded to it. Such data was considered to for simplicity of studying the quality of the created algorithms. However, the developed algorithms can be easily applied for the case when each program was run several times. There are at least $2$ ways to apply the described algorithms in this case:
\begin{enumerate}
    \item Form a $PC$ matrix as described in Section 2. In this case, a single program will correspond to a group of strings. Then the algorithms proposed in this article in Section 5 are applied;

    \item In case of several program executions, you can first make predictions based on the history – predictions of the program execution time within a single computer installation \cite{Gibbons_1997}, \cite{Kapadia_1999}, \cite{Li_2004}, \cite{Mohr_2003}. This way, you can fill in the empty entries in the columns. Then the algorithms proposed in this article in Section 5 are applied to fill the remaining empty entries.
\end{enumerate}

\section{Conclusion}
This article describes algorithms for predicting the Programs execution time on the federation’s computer installations that meet the requirements of the virtual infrastructure. Existing algorithms for predicting program execution time use the history of executions to make a prediction. The main drawback of existing algorithms is that the algorithms work only within a single computer installation, i.e. to predict the execution time on a certain computer installation, you need more than one start of the program on this computer installation – that is, you can’t do without the history.

In this paper, a new approach that allows to predict the Program execution time on a certain computer installation was proposed, even when it was not executed on it. Two algorithms were constructed and analyzed: an algorithm based on computer installations grouping based on the Pearson correlation coefficient (for cases when the Program was executed on almost all federates in the federation) and an algorithm based on matrix decomposition technique, which allows to obtain vector representations (embeddings) of the Program and computer installations (for cases when the Program was executed only on a small subset of federates). Matrix decomposition, which was used to predict the Programs execution times, is actively used in the field of recommendation systems, for example, SVD decomposition \cite{SVD}, ALS \cite{ALS_DECOMPOSITION} and tensor decomposition \cite{TENSOR_DECOMPOSITION}. In the article, for the first time, it is proposed to use this technique for the problem of predicting the programs execution time on an set of computer installations.

In addition, an ensemble of algorithms consisting of Ridge regression, grouping computer installations based on Pearson correlation coefficient, and matrix decomposition was proposed. It is shown that the ensemble of algorithms is more resistant to outliers than other algorithms and gives the best results on dense matrices.

Embeddings of programs and computer installations were obtained as a result of matrix decomposition of the matrix "Programs-Computers", where value in each entry is the program execution time on the computer installation. This approach to obtaining embeddings is widely used in many areas, for example, in recommendation systems \cite{RECOM_SYS}, in thematic modeling \cite{TOPIC_MOD}. This approach was successfully applied in solving the problem. The approach proved to be universal, as similar results were obtained for MPI benchmarks and OpenMP benchmarks (see experiments in Section 6).

It is important to emphasize that the proposed approach to predicting program execution time requires a minimal set of data about the program, which is usually available on all modern computer installations. Another important advantage of the proposed approach is that a result of matrix decomposition is the embeddings of Programs and computer installations of dimension 1. This fact allows one set up the total order as on a set computer installations as on a set of programs what significantly help to properly select the computer installation with effective execution time.
 
The final remark is as a hypothesis we state that execution time prediction techniques proposed in this article one can apply not only to MPI programs.

\end{document}